\documentclass[1p]{elsarticle}
\usepackage[draft]{hyperref}
\usepackage{graphicx,amssymb,amsmath,multirow}
\journal{Astronomy and Computing}
\biboptions{authoryear}

\begin{document}
\newcommand{\be}{\begin{equation}}
\newcommand{\ee}{\end{equation}}
\newcommand{\bq}{\begin{eqnarray}}
\newcommand{\eq}{\end{eqnarray}}
\begin{frontmatter}

\title{Abelian-Higgs Cosmic String Evolution with CUDA}
\author[inst1,inst2,inst3]{J. R. C. C. C. Correia}\ead{Jose.Correia@astro.up.pt}
\author[inst1,inst2]{C. J. A. P. Martins\corref{cor1}}\ead{Carlos.Martins@astro.up.pt}
\address[inst1]{Centro de Astrof\'{\i}sica da Universidade do Porto, Rua das Estrelas, 4150-762 Porto, Portugal}
\address[inst2]{Instituto de Astrof\'{\i}sica e Ci\^encias do Espa\c co, CAUP, Rua das Estrelas, 4150-762 Porto, Portugal}
\address[inst3]{Faculdade de Ci\^encias, Universidade do Porto, Rua do Campo Alegre 687, 4169-007 Porto, Portugal}
\cortext[cor1]{Corresponding author}

\begin{abstract}
Topological defects form at cosmological phase transitions by the Kibble mechanism, with cosmic strings---one-dimensional defects---being the most studied example. A rigorous analysis of their astrophysical consequences is limited by the availability of accurate numerical simulations, and therefore by hardware resources and computation time. Improving the speed and efficiency of existing codes is therefore important. All current cosmic string simulations were performed on Central Processing Units. In previous work we presented a General Purpose Graphics Processing Unit implementation of the evolution of cosmological domain wall networks. Here we discuss an analogous implementation for local Abelian-Higgs string networks. We discuss the implementation algorithm (including the discretization used and how to calculate network averaged quantities) and then showcase its performance and current bottlenecks. We validate the code by directly comparing our results for the canonical scaling properties of the networks in the radiation and matter eras with those in the literature, finding very good agreement. We finally highlight possible directions for improving the scalability of the code.
\end{abstract}


\end{frontmatter}


\section{\label{intr}Introduction}
A generic cosmological prediction of many theories beyond the Standard Model is the formation of objects known as topological defects, by means of the Kibble mechanism \citep{Kibble:1976sj}. Since properties of these objects and their astrophysical consequences are intrinsically linked to the symmetry breaking patterns which produce them, one can think of them as fossil relics of the physical conditions in the early Universe.

Recent constraints on these objects using cosmic microwave background and gravitational wave data \citep{PlanckDefects,LIGODefects} are mainly limited by the existence of accurate high-resolution simulations of defect networks with a large dynamic range, as well as full sky maps of the backgrounds produced by these networks: one contemporary example is the search for cosmic strings in cosmic microwave background maps using the Kaiser-Stebbins effect (for which one would ideally like to have thousands of statistically independent template sky maps, generated from a similar number of independent simulations, as opposed to current analyses relying on one or a few simulations), and analogous searches will undoubtedly be done with gravitational wave maps in the near future. The approximations currently being used to mitigate the absence of such data clearly introduce systematic uncertainties that are comparable to the quoted statistical uncertainties. This problem is even more severe for next-generation facilities such as CORE \citep{CORE} or LISA \citep{LISA}. On the other hand, analytic studies of realistic defect networks can---at least in principle---include enough degrees of freedom to explicitly model the relevant dynamical properties of these networks, but they will also be bottlenecked by the lack of high-resolution simulations, since these simulations are necessary to quantitatively calibrate the models. Resolving this issue by traditional means would imply unrealistic hardware/compute time needs.

To alleviate this problem, one can attempt to exploit differing hardware architectures with the onus of optimisation falling to the developers of the tool in question. In the literature there are several examples of defect simulations optimised for Central Processing Units (CPUs), either assuming shared or distributed memory architectures. Several examples of Goto-Nambu cosmic string simulations can be found in the literature \citep{BB,AS,FRAC,VVO,Blanco}, while examples of field theory simulations of the two simplest types of defect are the WALLS code of \citet{Rybak1} for domain walls---also optimised for Intel Xeon Phi co-processors, as summarised in \citet{Intel}---and the cosmic string evolution codes of \citet{Bevis:2006mj}. More recently, field theory defect simulations have also been developed for monopoles \citep{Monopoles}, global strings \citep{GlobalH,GlobalD}, Type I strings \citep{Type1}, semilocal strings \citep{Semilocals}, dual-higgsed strings \citep{Bevis:2008hg,Lizarraga:2016hpd} and hybrid defect simulations \citep{HindNab,McGraw,Hindmarsh:2018zch}. Last but not least, fully general relativistic treatments have also been recently introduced \citep{Helfer,GlobalD}.

Simulations which use Graphics Processing Units are far more scarce, with the only reported instance so far being \citet{PhysRevE.96.043310} by the authors, for domain wall networks. This paper is a continuation of our previous study, which seeks to simulate local Abelian-Higgs strings. We first introduce the algorithm used to simulate field theory cosmic strings (including the discretization used and how to calculate the relevant network averaged quantities), then showcase the performance of the implementation, and finally validate it by directly comparing with results found in the literature. We conclude by highlighting tentative directions to augment the scalability of this code. Meanwhile, the early results of the cosmological exploitation of this code can be found in \citet{Correia:2019bdl}.

\section{\label{disc}Discretization scheme}

A $U(1)$ local Abelian-Higgs string corresponds to a topological soliton that arises as a solution to the equations of motion of the Lagrangian
\begin{equation}
    \mathcal{L} = |D_{\mu}\phi|^2 -\frac{1}{4} F^{\mu\nu}F_{\mu\nu} - \frac{\lambda}{4}(|\phi|^2 -\sigma^2)^2\,,
\end{equation}
where $\phi$ is a complex scalar field, $F_{\mu\nu} = \partial_{\mu}A_{\nu} - \partial_{\nu}A_{\mu}$ a gauge field strength ($A_{\mu}$ corresponds to the gauge field), $D_\mu = \partial_{\mu} -ieA_{\mu}$ denotes a covariant derivative and $\lambda$ and $e$ are two constants which set the values of the scalar and vector masses, respectively $m_\phi = \sqrt{\lambda} \sigma$ and $m_v = e\sigma$.

We follow the same discretization procedure as \citet{Bevis:2006mj}, which requires first writing the discrete Lagrangian,
    \begin{equation}
    \begin{aligned}
    \mathcal{L} &= \frac{1}{2e^2 (\eta) a^2 (\eta) \Delta x^2} \bigg( \sum_i (E_{i}^{x,\eta-1/2})^2 \\ 
    &- \frac{1}{2 \Delta x^2} \sum_i \sum_j \bigg[ 1-\cos( \Xi_{ij}^{x} ) \bigg] \bigg) + |\Pi^{x,\eta}|^2 \\
    &- \frac{1}{\Delta x^2} \sum_i |e^{-i A_{i}^{x,\eta-1/2}} \phi^{x+k_i,\eta} - \phi^{x,\eta}|^2  \\
    &- \frac{1}{4} a^2 (\eta) \lambda(\eta) (|\phi^{x,\eta}|^2 -\sigma^2 )^2
    \end{aligned}
    \end{equation}
where $a = a(\eta)$ is the conformal factor for a Friedmann-Robertson-Walker space-time, $\phi$ and $A_i$ represent the complex scalar field and the spatial components of the gauge field at conformal time-step $\eta$ and  $\Pi$ and $E_i$ are conjugates of the aforementioned fields at half-steps ($\dot{\phi}$ and $\dot{A}_i$, respectively). Note that vector fields are rescaled as $A_i^x \rightarrow e \Delta x A_i^x $ and $ E_i^x \rightarrow e \Delta \eta E_i^x $. Both scalar fields reside at lattice sites $i,j,k \rightarrow x$, and the vector fields at half-sites (for convenience however site $A^{x+k_i/2}_i$, where $k_i$ is some unit vector, is written $A^{x}_i$).

The action of the gauge field on the lattice is then defined in terms of link variables, as for instance $\exp(-iA^{x}_i)$. These are a standard prescription of lattice gauge theory to construct gauge invariant quantities on a lattice, see \citep{Wilson:1974sk}. They can then be used to define a plaquette operator 
\begin{equation}
\exp(iA_{i}^{x})\exp(iA_{j}^{x+k_i})\exp(-iA_{i}^{x+k_j})\exp(-iA_{j}^{x}) = \exp(i\Xi_{ij})\,,
\end{equation}
where $\Xi_{ij} / \Delta x^2$ represents the discrete version of $F_{ij}$. The real part of the plaquette operator in the small lattice spacing limit, along with the small-angle approximation ($1-\cos(x) \rightarrow 0.5x^2 $) defines the discretized version of $F_{ij}F_{ij}$. The gauge covariant derivative is written as $\exp(-i A_{i}^{x,\eta-1/2}) \phi^{x+k_i,\eta} - \phi^{x,\eta}$ . Finally $\sigma$ sets the vacuum expectation value, $a_{\eta}$, $e_{\eta}$, $\lambda_{\eta}$ respectively denote the scale factor, the gauge and scalar couplings at time $\eta$ (to be defined below), and $\Delta x$ and $\Delta \eta$ are the lattice spacing and time-step values.

Through variational principles, much as can be found in \citet{Bevis:2006mj}, one can write the equations of motion as a staggered leap-frog scheme
\begin{equation} \label{eq:up_pi}
    \begin{aligned}
    \Pi^{x, \eta +1/2} &=& { \bigg( \frac{a_{\eta-1/2}}{a_{\eta+1/2}} \bigg) }^2 \Pi^{x, \eta-1/2}-\Delta \eta {\bigg( \frac{a_{\eta}^2}{a_{\eta+1/2}} \bigg)}^2 \frac{\lambda_{\eta}}{2}( |\phi^{x,\eta}|^2 -\sigma^2) \phi^{x,\eta}
                        \\ & &+\frac{\Delta \eta}{(\Delta x)^2} { \bigg( \frac{a_\eta}{a_{\eta+1/2}} \bigg) }^2 \sum_j \bigg( \phi^{x+k_{j}} \exp(iA^{x,\eta}_j) -2 \phi^{x,\eta} + \phi^{x-k_{j},\eta} \exp(-iA^{x-k_{j},\eta}_j) \bigg)
    \end{aligned}
\end{equation}
\begin{equation} \label{eq:up_E}
    \begin{aligned}
    E_{i}^{x,\eta+1/2} &=& { \bigg( \frac{e_{\eta+1/2}}{e_{\eta-1/2}} \bigg) }^2 E_{i}^{x, \eta-1/2} + 2 \Delta \eta a_{\eta}^2  e_{\eta+1/2}^{2} {\rm Im} \bigg[ (\phi^{x+k_i,\eta})^{*}\exp(-i A^{x, \eta}) \phi^{x,\eta} \bigg]
        \\ & &- \frac{\Delta \eta}{(\Delta x)^2} \bigg( \frac{e_{\eta+1/2}}{e_{\eta}} \bigg)^2 \sum_{j \neq i} \bigg[ \sin(\Xi_{ij}^{x}) - \sin(\Xi_{ij}^{x-k_j}) \bigg] 
    \end{aligned}
\end{equation}
\begin{equation} \label{eq:up_phi}
\phi^{x, \eta+1} = \phi^{x,\eta} + \Delta \eta \Pi^{x, \eta+1/2}
\end{equation}
\begin{equation} \label{eq:up_A}
A_{i}^{x,\eta+1} = A_{i}^{x,\eta} +\Delta \eta E_{i}^{x,\eta+1/2}, 
\end{equation}
which tells us how to update field variables at each time-step. The simulations are evolved until half-a-light-crossing time, since due to the periodic boundary conditions, evolving any further would affect the dynamics (all the defects would effectively be inside the same horizon which would not adequately mimic a network in an expanding universe).

There is a further subtlety in most field theory defect simulations: since the physical defect width is constant, it will shrink in comoving coordinates. This means that the true equations of motion result in strings that eventually fall through the lattice and can no longer be resolved. A way to bypass this problem is to fix the comoving width as originally done in \citet{PRS} or to first apply a comoving core growth period \citep{Bevis:2006mj}, such that by the end of the simulation defects can still be resolved. Since the defect width is inversely proportional to the scalar and vector masses, this means that one must change the way $e_\eta$ and $\lambda_\eta$ behave as
\begin{align}
	e_\eta = e_0 a^{s-1}_\eta && \lambda_\eta = \lambda_0 a^{2(s-1)}_\eta
\end{align}
where $s$ is a parameter such that $s=0$ will force constant comoving core width and $s=1$ recovers the original equations of motion (negative values imply core growth). As long as these modified constants are used in the action (and consequently in the discrete version of the equations of motion above), then this parameter can be controlled throughout the simulation to either have defects with fixed comoving width or with growing width up to some specified timestep, and then physical width thereafter.

For our simulations scalar and vector masses are made equal by choosing $e_0=1$ and $\lambda_0=2$, and in what follows we will present results both with $s=0$ and with $s=1$. For the initial conditions, we choose to mimic a field configuration after a phase transition, that at the same time obeys the discretized form of Gauss's law. As such, the scalar field $\phi$ is set to have a random phase, its norm is set to unity (given that $\sigma=1$) and all other field variables are set to zero. For the random phase, the library cuRAND \citep{curand} is used. 

In order to validate the simulations, two diagnostics are calculated, a mean string separation and a weighted mean squared velocity (taken from local gradient and rate of change of the scalar field),
\begin{align} \label{defXiV2}
    \xi_{\mathcal{L}} = \sqrt{\frac{-\mu}{\bar{\mathcal{L}}}} && \langle v^2 \rangle_{\mathcal{X}} = \frac{2R}{1+R} 
\end{align}
 where
\begin{equation} \label{defR}
  R = \frac{\int |\dot{\phi}|^2 \mathcal{X} d^3 x}{\int |D\phi|^2 \mathcal{X} d^3 x}.
 \end{equation} 
 and $\mathcal{X}$ is a weight function.
The first estimator is taken from \citet{Bevis:2006mj} where it was shown that the discrete Lagrangian density peaks negatively at the strings (and therefore its mean, $\bar{\mathcal{L}}$ can be used for mean string separation estimation), and the second matches the corrected definition presented in \citet{Hindmarsh:2017qff}. In what follows we will use two different weighting functions---either the potential or the Lagrangian---in order to establish a comparison.

As a cross-check we use a further mean string separation estimator, $\xi_W$, where the total length of string is computed by finding all lattice plaquettes pierced by a string. The way to do this is to consider a gauge invariant winding around a plaquette,
\begin{equation}
	W_{ij} = (Y_{i}^{x} + Y_{j}^{x+i} - Y_{i}^{x+j} - Y_{j}^{x})/2\pi\,,
\end{equation}
where $Y_{i,x} = [(\phi^{x+k_i})_{arg} -(\phi^{x})_{arg}+A_{i}^{x}]_\pi -A_{i}^{x}$, as introduced in \citet{Kajantie:1998bg}. Note that the argument of the scalar field is assumed to be between $[-\pi,\pi]$ and the term in $[...]_{\pi}$ has $\pi$ factors added or subtracted in order to bind the result to the interval $[-\pi,\pi]$. If $W_{ij}$ is different from zero then a string is present, and a length of $\Delta x$ is associated with each winding found. Note that we also multiply by a factor of $\pi/6$ to account for the Manhattan effect \citep{PhysRevD.58.103501}.

A further cross-check can be done through an estimator of \citep{Hindmarsh:2017qff} for the mean velocity squared, where the mean weighted pressure and density are used to compute the equation of state parameter of the strings and from it the velocity,

\begin{equation} \label{defEoS}
\langle v^2 \rangle_\omega = \frac{1}{2} \bigg(1+3\frac{p_{\mathcal{L}}}{\rho_\mathcal{L}} \bigg)
\end{equation}
Given that in \citet{Hindmarsh:2017qff} this estimator has been shown to be in better agreement with the velocity of a standing wave string, this serves as a baseline for the comparison between estimators.

\section{\label{impl}Implementation and performance}

We now describe how our implementation utilises an application programming interface named Compute Unified Device Architecture (CUDA, by NVIDIA Corporation) to evolve a network of Abelian-Higgs cosmic strings. The development and all benchmarks were done on an NVIDIA Quadro P5000, with 2560 CUDA cores, a core clock of 1607 MHz and 16384 MB of memory, clocked at 1126 MHz.

One of the main roles of CUDA is to provide a way to abstract some details of the underlying hardware, while allowing some degree of optimisation. A relevant example in the present case is that even though GPU's are made of Streaming Multiprocessors (each made up of several cores which execute instructions in a 32-way lane---called a warp---in Single Instruction Multiple Data fashion), we will not explicitly distribute threads across different Streaming Multiprocessors, but spawn a number of threads equal to $N^2$ in an $N^3$ simulation box (the reason for this will become apparent in two paragraphs), subdivided into groups of threads called thread blocks. Multiple thread blocks will be resident at each Streaming Multiprocessor, assigned automatically without our intervention.

In CUDA, applications are subdivided into data parallel functions named kernels. In our application there are three kernels that evolve the field configurations at every time-step: the first one corresponds to Eq. \ref{eq:up_pi}, the second to Eq. \ref{eq:up_E} and the third one to Eqs. \ref{eq:up_phi}--\ref{eq:up_A}. These will be denoted stepA, stepB and stepC respectively. There are also kernels associated with computing useful quantities such as the mean string separations or the velocities. Since these kernels implement essentially finite differences (and often these are memory-bound) one must exploit the memory hierarchy of a GPU. The abstract memory model of CUDA describes myriad types of memory and the ones relevant for the next paragraph include: global memory (which corresponds to video memory), shared (a fast-on-chip memory available to groups of threads, known as thread blocks) and registers (per-thread memory, even faster on-chip memory).

\begin{figure}
\begin{center}
\includegraphics[width=8cm]{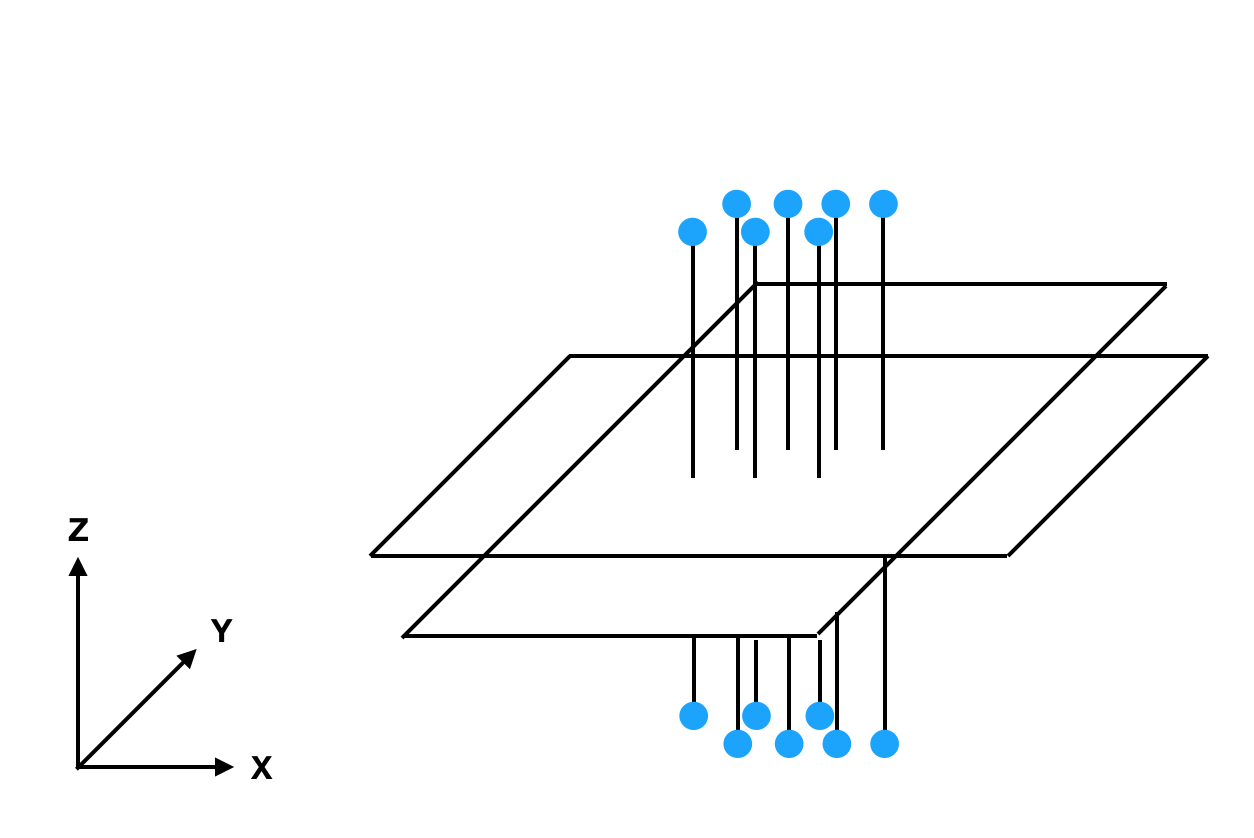}
\caption{\label{fig1}Schematic representation of the stepA kernel (which updates the conjugate momentum $\Pi$): the tile in the middle represents a 2D shared memory tile where current values in the z-direction (site $k$) are loaded together with halos, and register values (blue pinheads) hold field values directly above and below ($k-1$ and $k+1$).}
\end{center}
\end{figure}

For the first two kernels one loads relevant field quantities from global memory at zero height ($k=0$) to shared memory. We denote these 2D (oriented along X and Y) chunks of shared memory as tiles. Note that one would naively expect these tiles to have size equal to the number of threads along the $x$-direction times the number of threads in the $y$-direction in a given thread block, however, as can be seen from the discretization scheme,  there are terms which involve using field quantities in positions $x+k_j$, $x-k_j$ where $j=1,2,3$. As such, the very frontiers of each tile require values from neighbouring tiles. Given that there is no communication between thread blocks, we must pad our XY-tiles by 2 along each direction and load appropriate boundary terms to these padding regions (commonly known as ghost cells or halos) prior to any actual computation. Afterwards we simply stream through the z-direction \citep{Zhang,Micikevicius,5645463,5470394}. The main advantage of
  doing so is to enable software pre-fetching: load only the field at the next z-position (into registers) and when streaming up the z-direction this  value is loaded into the current shared memory tile. Similarly the previously current shared memory tile is loaded into the bottom registers (see Fig. \ref{fig1} for a schematic representation). In stepB, and in the kernels which calculate average network quantities, instead of temporary variables above and below, we use shared memory tiles for the top and bottom (complete with halos). There are two reasons to do so: for convenience (some calculations may require values on the top/bottom tile's halos) and to reduce register over-use. Note that if the amount of necessary registers exceeds what can be provided by on-chip memory, the compiler will 'spill' some of these variables to the slower global memory. This is known as register spilling.

There is another advantage to loading field values into shared memory tiles and/or registers. The field variables are given by the aligned vector types defined in CUDA (float2 and float4) and while the vector loads ensure coalesced\footnote{Coalesced memory accesses are defined as accesses where multiple reads or writes into one single transaction. In general this requires that such reads/writes are not sparse, misaligned or non-sequential. In NVIDIA GPU's every successive 128 bytes can be loaded by 32 threads (denoted a warp) in one single transaction ($32 \times 4$ bytes then yields 128 bytes).} memory reads, some computations which require specific components of each field would cause un-coalesced reads. This bottleneck is circumvented by using shared memory and registers. The third kernel is more straightforward, since software pre-fetching cannot be implemented. It simply reads the fields and their conjugates from global memory and writes the updated field values again.

\begin{table}
\begin{center}
\caption{The effective Global Load and Store bandwidth (in units of GB/s), the number of Floating Point Operations per second (in Teraflops) and the achieved occupancy, for a $256^3$ simulation in the radiation era and for constant comoving width.}
\label{table1}
\begin{tabular}{| c | c c | c |}
\hline
 Kernel & GLS (GB/s) & TFLOPs & Occupancy (\%) \\
 \hline                                                                                                                                     
 stepA		& $245.92$ & $1.59$	 	 & $48.0\%$		\\			     
 stepB		& $271.60$ & $0.80$	 	 & $47.8\%$		\\			    
 stepC		& $264.83$ & $0.04$ 	 	 & $90.8\%$		\\
\hline
 VelRVLag   & $212.88$ & $1.68$ 	 	 & $48.3\%$ 		\\	
 VelRWLag   & $217.09$ & $1.67$ 	 	 & $48.3\%$ 		\\ 
 VelEoSLag  & $193.70$ & $1.80$ 	 	 & $48.3\%$ 		\\
\hline 
Winding    & $133.72$ & $1.26$ 	 	 & $48.3\%$ 		\\
 \hline
\end{tabular}
\end{center}
\end{table}

The three kernels are limited by memory bandwidth, in particular when reading from global memory, as indicated by the NVIDIA Visual Profiler. As such, the most relevant performance metric is the effective bandwidth (bytes loaded and stored from/into global memory per second) and how it compares to the peak bandwidth of the GDDR5X memory present in the test-bench graphics card. The average bandwidth reached for each kernel (together with additional metrics) can be found in Table \ref{table1}, for box size $256^3$, in the radiation era and for constant comoving width. It is seen that that we are close to peak bandwidth (288.5 GB/s). Additionally the peak single precision is of 8.876 TFLOP/s on this particular card. Both peak throughput and bandwidth are expected theoretical values, reported by the NVIDIA Visual Profiler (ie. we did not perform measurements for peak bandwidth/throughput using custom kernels).

An important detail is the chosen size of thread block, in particular for the first two kernels. In general, kernels which perform finite difference methods (such as those of a 7-point stencil) prefer a larger thread block size in order to mitigate the performance hit from loading tile halos. However, in our case, due to the data-reuse pattern above one must also consider if the thread-block size will not result in register/shared memory overuse. With the help of the online CUDA Occupancy Calculator \citep{occup}, the thread-block size that seemed to yield best performance was (32,4) at $256^3$ box size. The main limiting factor for the occupancy per Streaming Multiprocessor (the ratio of warps being executed upon each Streaming Multiprocessor to the theoretical maximum number of warps per Streaming Multiprocessor) seems to be register pressure, as shown by the NVIDIA Visual Profiler. In all cases the occupancy is large enough that increasing it might not yield better performance: as previously stated, the Visual profiler does not indicate latency as the main performance bottleneck.

The kernels which calculate the mean string separation $\xi_{\mathcal{L}}$ (recall Eq. \ref{defXiV2}) and each of three mean velocity squared estimators $\langle v^2 \rangle_\mathcal{V}$, $\langle v^2 \rangle_\mathcal{W}$ or $\langle v^2\rangle_\omega$ (hereinafter named VelRVLag, VelRWLag and VelEoSLag, respectively), as well as the kernel which computes the winding (named Winding) operate by using the memory pattern described above to load data into shared memory tiles. They then compute the Lagrangian, and either the numerator and denominator of $R$, or the ratio between $\rho_\mathcal{L}$ and $p_{\mathcal{L}}$ for each thread (cf. Eqs. \ref{defR} and Eq. \ref{defEoS}, respectively). Each result is stored in a register, and the CUDA Unbound library \citep{CUB} is finally used to compute a thread block-wide sum. Since each block computes a partial sum, we then transfer these back to the host and after summing we write to disk.

The partial sums are calculated on the GPU in order to avoid becoming IO-bound (PCI-E buses could be easily saturated by transferring the values of each field variable to the host). The three velocity estimator kernels are bottlenecked by both compute and memory requirements, and end up having reasonably similar performances. The memory requirements are in part explained by the excessive register spilling that occurs (this can be avoided by not limiting the maximum number of registers to 64 per thread with the compiler option \texttt{--maxrregcount}, but the side-effect is that it significantly reduces the occupancy, and this heavily impacts performance). The impact of spilling is mitigated by turning on the compiler flag \texttt{--Xptxas dlcm=ca} which caches these spills in L1. Improving the compute part however is more challenging: many of the compiler flags which attempt utilisation of hardware intrinsics, or reduce the precision of certain operations often affect the quality of the diagnostics, either changing the asymptotic quantities themselves or increasing uncertainties.

Still there is one simple optimisation that reduces runtimes: avoid executing this kernel at every timestep. In other words, we calculate the diagnostic quantities every $n$ timesteps only (hereinafter we take $n=5$), and reduce statistical uncertainties by doing multiple runs. This effectively reduces the time spent in the calculation kernels, as can be seen in Table \ref{table2} for an example run where all estimators are run. Note however that in typical production runs one will select only one of the of the velocity estimators and optionally the Winding estimator. The total run time will therefore depend on what diagnostics one chooses to output, and how often this is done.

\begin{table}
\begin{center}
\caption{Total elapsed times, in seconds, on the three evolution kernels plus estimator kernels (which calculate averaged network quantities) of one $256^3$ and one $512^3$ run. The total time is computed by taking the average runtime and multiplying by the number of times a kernel is executed in a single run, i.e. $\langle Time \rangle \times n_{\rm calls}$. The first three kernels are executed every timestep, while the others are executed only every 5 timesteps.}
\label{table2}
\begin{tabular}{| c | c | c |}
\hline
Kernel & $256^3$   & $512^3$  \\
\hline
 stepA     & 2.29   & 36.86 \\
 stepB     & 3.10   & 50.02 \\
 stepC     & 2.89   & 46.30 \\
\hline
 VelRVLag  & 0.57   & 8.38  \\
 VelRWLag  & 0.56   & 8.45  \\
 VelEoSLag & 0.64   & 8.88  \\
\hline
 Winding  & 0.87   & 11.92 \\
 \hline
\end{tabular}
\end{center}
\end{table}

One final remark about the time spent in Input/Output operations (transferring partial sums, computing the final sum on the host, cf. Table \ref{table2}) is that we can speed up the simulation further by overlapping compute on the GPU with the aforementioned operations: however, how much can be gained in terms of speed will also depend on how often we choose to calculate useful quantities. For now, given that a reduced number of calls to estimator kernels diminishes the need for such an optimisation and that we can venture into multi-GPU territory, we keep everything non-overlapped.

The main motivation for writing GPU-optimised applications is based on the higher theoretical bandwidth and throughput ceilings. Based on typical figures for current high-end top-of-the-line multicore CPUs, memory and GPUs, one can stipulate a speed-up of one order of magnitude for bandwidth or compute-bound applications \citep{CPguide} assuming of course both applications are fully multi-threaded and optimised (and thus reach close-to-peak throughput and bandwidth). Of course this will depend on the underlying hardware where each simulation is executed and on the optimisation applied, so the precise number may vary. In our previous work on domain walls \citep{PhysRevE.96.043310} a speed-up of about two orders of magnitude was found in a typical desktop computer when comparing to a single-thread implementation. In the present case we don't have a CPU code that enables a direct comparison (and the published work of other authors does not provide useful benchmarks), but we could conservatively estimate a speed-up of at least one order of magnitude.

We can also compare the performance in time to full evolution multiplied by the number of processors per number of sites (in either gpu-sec/site vs core-sec/site) for the evolution update and winding outputs with the cosmic string simulation of \citep{Bevis:2006mj,Hindmarsh:2017qff} (hereby referred to as Lattice Abelian-Higgs - LAH) measured in the Monte Rosa supercomputer at $4096^3$ box size with $32768$ cores \citep{Hind:email}. Note that before such a comparison can be made there are some caveats: \citet{Hind:email} remarked that their simulation is not too optimised. While we were provided the performance of the winding update, note that most of time spent on windings is due to output (writing to disk), not due to the computation of windings (in contrast with our case, as we sum the windings and output the mean string separation estimator only). As such it is not entirely correct to compare the winding performance directly with ours. LAH for the evolution update (that is, the analogous computation to our stepA + stepB + stepC), for 10903 timesteps, $4096^3$ box size and for the computation of winding (for 1300 of the timesteps) has the following performance figures: $8\cdot 10^{-7}$ core-sec/site and $1.3\cdot10^{-6}$ core-sec/site, respectively. Evolving a $512^3$ lattice from start to finish (stepA + stepB + stepC for 1280 timesteps) reveals a performance of $7.75\cdot10^{-10}$ gpu*sec/site. Compared with the evolution figures from LAH, our simulation therefore spends about three orders of magnitude less time updating fields on a given lattice site, showing that our estimate in the previous paragraph is indeed conservative. We present the figures for all of the other kernels in Table \ref{tableCompar}. Note that in general GPU cores are much slower than traditional processor cores, even though this last table seems to suggest they are only $2.5$ times slower. We are not sure of the reason for this behavior, but we may speculatively suggest that it reflects the different levels of optimization of the two codes. 

 \begin{table*}
    \begin{center}
	    \caption{The performance of each of our kernels given in gpu-sec/site. Note that in order to compare with the LAH performance, provided by \citet{Hind:email}, we present the performance of all of three update kernels together (stepA+B+C, computed by summing the time for each update kernel from Table \ref{table2} and then dividing by the number of sites). These numbers can be obtained from the times at $512^3$ in Table \ref{table2} by dividing by the number of calls of each kernel in a run (1280 for steps A and B and C, 256 for estimators) and by dividing over the size of the lattice $512^3$.}
    \label{tableCompar}
    \begin{tabular}{| c | c | c |}
    \hline
    Kernel &  Performance GPU-AH        & Performance LAH\\
           &  (gpu-sec/site at $512^3$) & (core-sec/site at $4096^3$)\\
    \hline
     stepA+B+C & $7.75 \cdot 10^{-10}$ & $8 \cdot 10^{-7}$ \\
    \hline
     VelRVLag     & $2.43\cdot10^{-10}$  & Not available \\
     VelRWLag     & $2.46\cdot10^{-10}$  & Not available \\
     VelEoSLag    & $2.58\cdot10^{-10}$  & Not available \\
    \hline
     Winding     & $3.47\cdot10^{-10}$ & $1.3 \cdot 10^{-6}$\\
     \hline
    \end{tabular}
    \end{center}
    \end{table*}

\section{\label{valid}Scaling Validation}

\begin{figure*}
\begin{center}
\includegraphics[width=8cm]{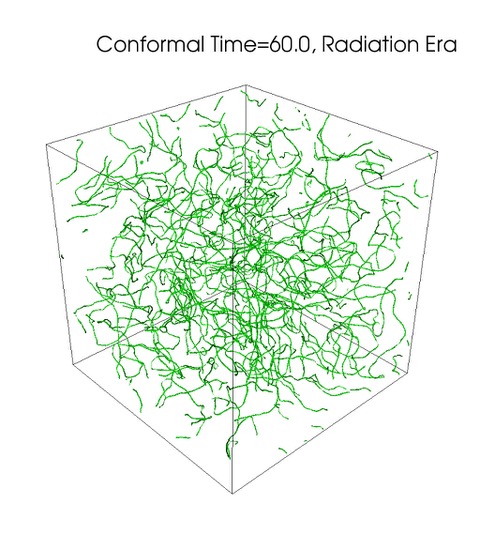}\includegraphics[width=8cm]{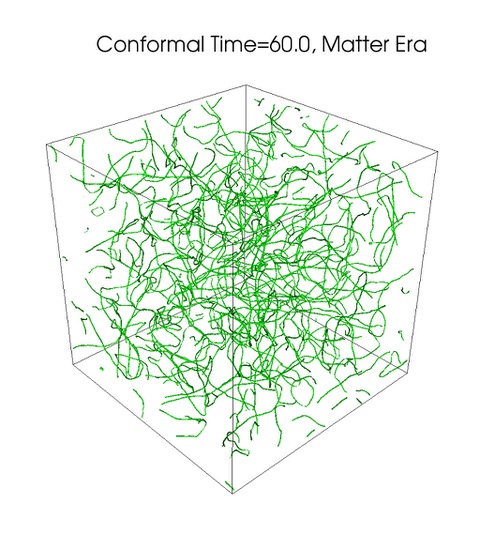}
\includegraphics[width=8cm]{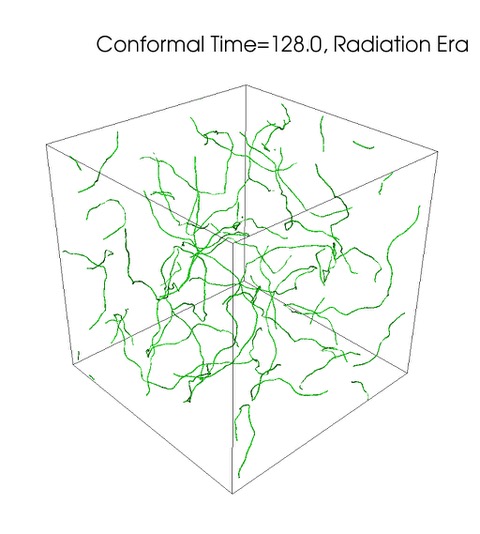}\includegraphics[width=8cm]{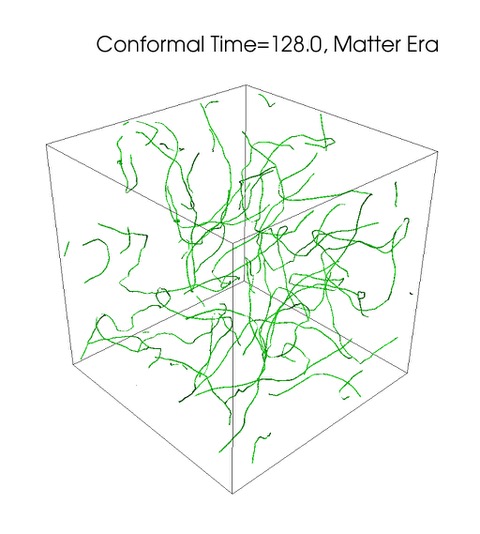}
\caption{\label{fig2}Isosurfaces of the absolute value of the complex scalar field with the value of 0.5, showing a network of Abelian-Higgs cosmic strings in the radiation and matter eras (left and right side panels respectively). All pictures are from simulations with box size $512^3$; the top panels correspond to timestep 60, while the bottom panels correspond to timestep 128.}
\end{center}
\end{figure*}

We have checked that the discretized form of Gauss's law is preserved to machine precision. Additionally, inspecting iso-surfaces of the scalar field provides visual confirmation that a network of strings is formed and evolves as expected---some examples can be seen in Fig. \ref{fig2}.

For the domain walls GPU code \citep{PhysRevE.96.043310} we had a serial version of the simulation which had been previously tested and validated \citep{PRS,Rybak1,Rybak2}, and could directly compare outputs. In the present strings case both the serial and parallel versions are completely new to the authors, so we will validate them by evaluating the asymptotic scaling values and comparing them with the results in the literature (which come from CPU codes). We have performed simulations in the two canonical cosmological epochs, the radiation and matter eras, for which the scale factor respectively evolves as $a(\eta)\propto\eta$ and $a(\eta)\propto\eta^2$. Snapshots of the simulations in the two eras can be seen in Figure \ref{fig2}. This comparison is summarised in Table \ref{table3}. The scaling quantities obtained in the present work are the averages of the velocity and slope of the mean string separation, in the dynamic range in which the networks have reached scaling. In each case we average 5 different runs (with random initial conditions) to obtain a statistical error.

\begin{table*}
\tiny
\begin{center}
\caption{Numerical results for asymptotic scaling quantities $\dot{\xi}$ (calculated using the Lagrangian or the winding estimator) and the three velocity estimators, for $s=0$ and $s=1$ (where applicable), from our simulations and from the literature. All quantities were measured in simulations with box sizes of $512^3$, except where otherwise noted. The {\it ext.} and {\it asy.} denote values that were extrapolated (rather than directly measured from the simulations) and inferred by visual inspection of Fig.9 of \citet{Hindmarsh:2017qff}; see the main text for further discussion of these.}
\label{table3}
\begin{tabular}{| c | c | c | c | c | c | c | c |}
\hline
Epoch & $s$ & $\dot{\xi}_{\mathcal{L}}$ & $\dot{\xi}_{W}$ & $\langle v^2 \rangle_{\mathcal{V}}$ & $\langle v^2 \rangle_{\mathcal{L}}$ & $\langle v^2 \rangle_\omega$ & Reference \\
\hline
Radiation & 1 & $0.33 \pm 0.02$ & - & - & - & - & \citet{Bevis:2006mj} \\
Radiation & 1 & - & - & - & - & $0.37\pm0.01$ (ext.) & \citet{Hindmarsh:2017qff}$@4096^3$ \\
Radiation & 1 & - & - & - & - & $0.30\pm0.01$ (asy.) & \citet{Hindmarsh:2017qff}$@4096^3$ \\
Radiation & 1 & $0.254\pm0.005$ & $0.265\pm0.005$ & - & - & - & \citet{Daverio:2015nva}$@4096^3$ \\
Radiation & 1 & $0.32\pm0.01$ & $0.32\pm0.03$ & $0.34\pm0.01$ & $0.31\pm0.01$ & $0.31\pm0.01$ & {\bf This work} \\
\hline
Radiation &0 & $0.31\pm0.02$ & - & -	 & - & - & \citet{Bevis:2006mj} \\
Radiation &0 & - & $0.26\pm0.02$ & - & - & - & \citet{Bevis:2010gj}$@1024^3$ \\
Radiation & 0 & $0.234\pm0.006$ & $0.244\pm0.005$ & - & - & - & \citet{Daverio:2015nva}$@4096^3$ \\
Radiation &0 & $0.30\pm0.02$ & $0.32\pm0.03$  & $0.34\pm0.01$ & $0.31\pm0.01$ & $0.32\pm0.01$ & {\bf This work} \\
\hline
Matter & 1 & - & - & - & - & $0.31\pm0.01$ (ext.) & \citet{Hindmarsh:2017qff}$@4096^3$ \\
Matter & 1 & - & - & - & - & $0.26\pm0.01$ (asy.) & \citet{Hindmarsh:2017qff}$@4096^3$ \\
Matter & 1 & $0.261\pm0.008$ & $0.277\pm0.008$ & - & - & - & \citet{Daverio:2015nva}$@4096^3$ \\
Matter & 0 & $0.30\pm0.01$ & - & - & - & - & \citet{Bevis:2006mj} \\
Matter & 0 & - & $0.28\pm0.01$ & - & - & - & \citet{Bevis:2010gj}$@1024^3$ \\
Matter & 0 & $0.235\pm0.008$ & $0.247\pm0.008$ & - & - & - & \citet{Daverio:2015nva}$@4096^3$ \\
Matter & 0 & $0.29\pm0.01$ & $0.29\pm0.02$   & $0.26\pm0.01$ & $0.27\pm0.01$ & $0.25\pm0.01$ & {\bf This work} \\
\hline
\end{tabular}
\end{center}
\end{table*}

Comparing directly our results for the slope of the mean string separation with the values of $\dot{\xi}_{\mathcal{L}}$ in \citet{Bevis:2006mj} we find excellent agreement for both matter and radiation era simulations. Our other length estimator, $\dot{\xi}_{W}$, is also in excellent agreement with the results of the first, but in mild disagreement (about 1.5 standard deviations, if one assumes Gaussian errors) with the value found in \citet{Bevis:2010gj}. The discrepancy increases if we compare with the even larger simulations of \citet{Daverio:2015nva}. As explained in \citet{Bevis:2010gj} this is a consequence of the fact that these works \citep{Bevis:2010gj,Daverio:2015nva} include an early period of cooling (introduced by modifying the equations of motion, which effectively changes the initial conditions), which is done with the goal of reaching scaling as quickly as possible. The combination of this choice and the extended dynamic range then leads to a slow drift in the $d\xi / d\eta$ value (changing the $\dot{\xi}$ from the $0.3$ value to about $0.28$ at $1024^3$ and then about $0.24$ at $4096^3$). Thus in this particular case the statistical disagreement is at about 3 standard deviations, but this also highlights the fact that these simulations also include systematic uncertainties due to the numerical implementation itself, which must be taken into account as improvements in hardware and software gradually reduce the statistical uncertainties.

\begin{figure*}
\begin{center}
\includegraphics[width=8cm]{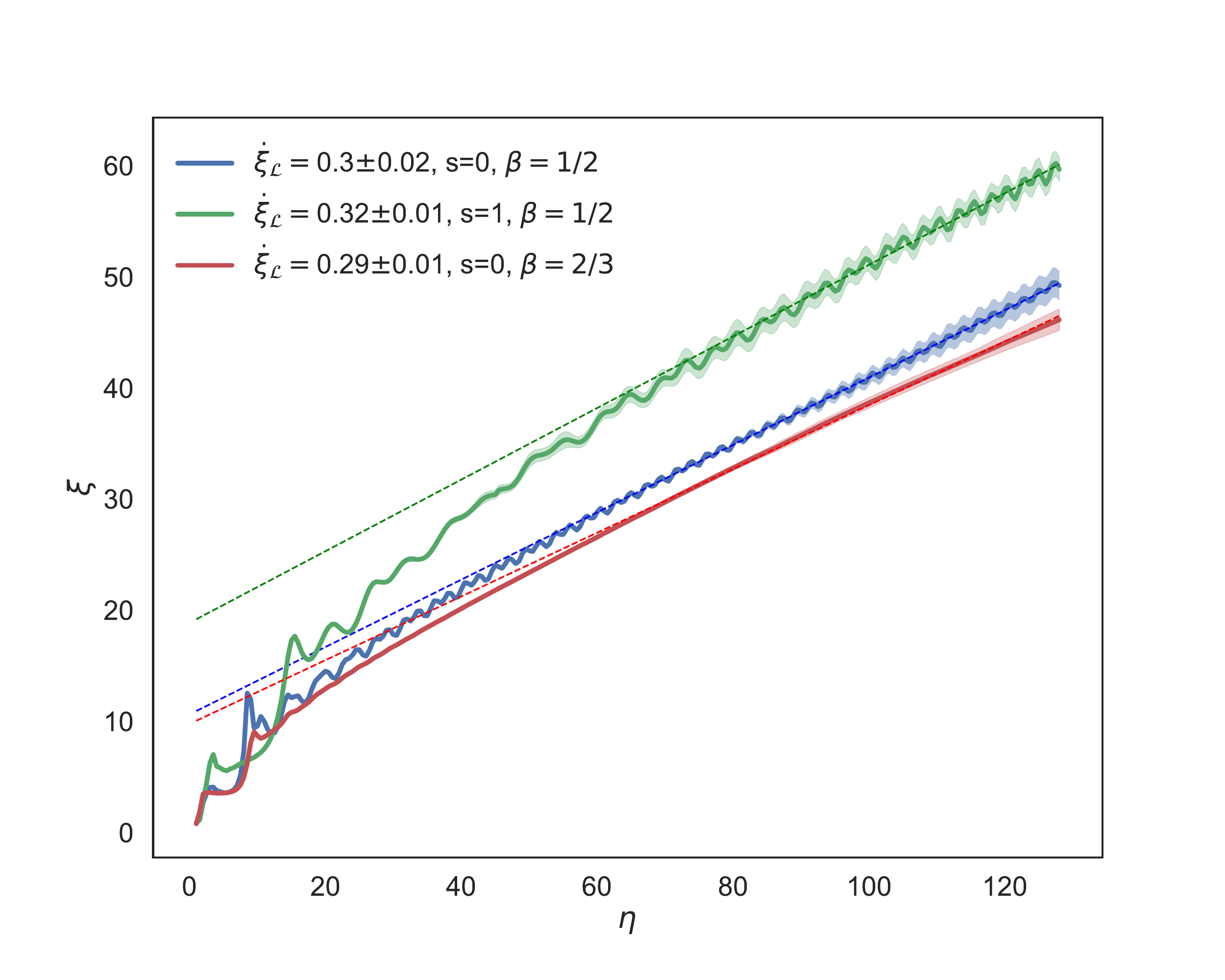}\includegraphics[width=8cm]{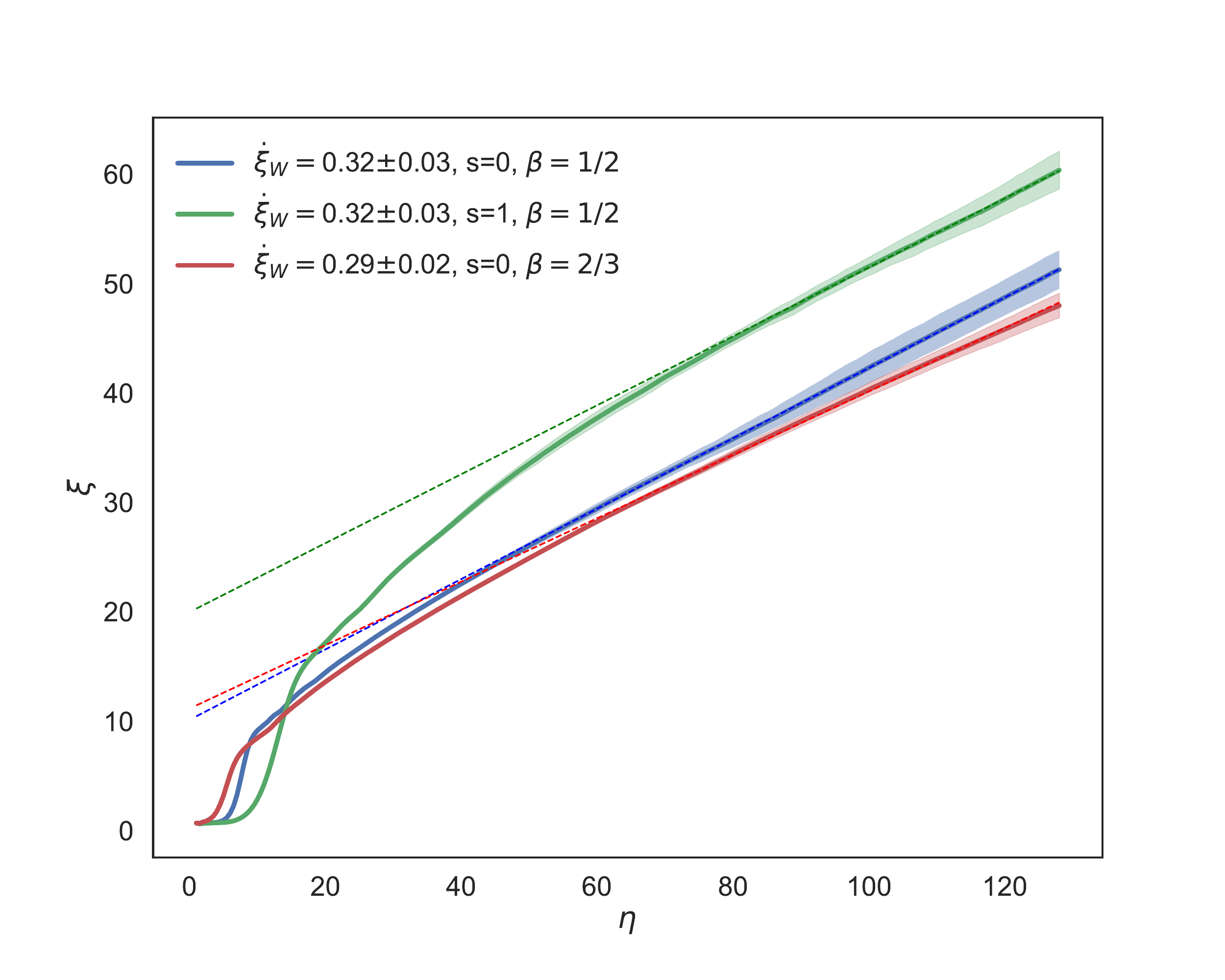}
\caption{\label{fig3}The evolution of the mean string separation ${\xi}_\mathcal{L}$ (left panel) and the winding based mean string separation ${\xi}_W$ (right panel) for $512^3$ runs, in the radiation era ($\beta=1/2$, blue lines without core growth and green lines with core growth) and in the matter era ($\beta=2/3$, red lines without core growth). The values of the mean string separation slopes, $\dot{\xi}$, inferred after the networks have reached scaling, are also added in the figure legends. These slopes are an average from the slopes of 5 different runs.}
\end{center}
\end{figure*}

Figure \ref{fig3} depicts the evolution of the Lagrangian-based mean string separation $\dot{\xi}_\mathcal{L}$ and the winding based mean string separation $\dot{\xi}_W$ for our $512^3$ runs, in both the radiation and matter eras. Qualitatively, the approach to scaling is clearly visible, and this is confirmed by the quantitative analysis described in the previous paragraph.

We note that in the case of the Lagrangian-based mean string separation some oscillations can be seen, signalling the presence of some radiation in the box. This is well understood from previous work with high-resolution field theory simulations of domain walls \citep{Rybak1,Rybak2}, which shows that the presence of this radiation does not prevent scaling. We note that it would be possible to artificially suppress this radiation by numerically implementing an ad hoc period of cooling, as is sometimes done in the literature. However, we have not done this: it is not necessary for our purposes (i.e., for validating the code through its diagnostics of scaling properties of the string networks). Indeed, since part of our goal is to demonstrate that scaling is reached, we should not use any evolution period which might artificially facilitate the approach to scaling. Moreover, previous work on domain walls shows that this would erase relevant information for the purpose of modelling of network evolution. Indeed, one can numerically separate the energy in defects from that in radiation, and analytically model the evolution of both: for domain walls this has been done in \citet{Rybak1,Rybak2}, and for strings early results demonstrating that this is possible can be found in \citet{Correia:2019bdl}.

As for the velocity estimator, the comparison has to be more qualitative since there are fewer measurements of velocities reported in previous field theory simulations. The most recent work is \citet{Hindmarsh:2017qff}, which only tabulates values obtained from extrapolating the results of their simulations to infinite string separation---a process whose physical meaning is not entirely clear. We do present these values in Table \ref{table3} (denoting them with {\it ext.}), but we also note that a more meaningful comparison is likely to be with the asymptotic values (denoted {\it asy.} in the table). The reason is simply that these asymptotic values were measured directly from simulations---indeed they can be visually read off from the top and bottom panels of figure 9 of \citet{Hindmarsh:2017qff}---while the others were extrapolated from the simulations with some additional assumptions.

\begin{figure*}
\begin{center}
\includegraphics[width=8cm]{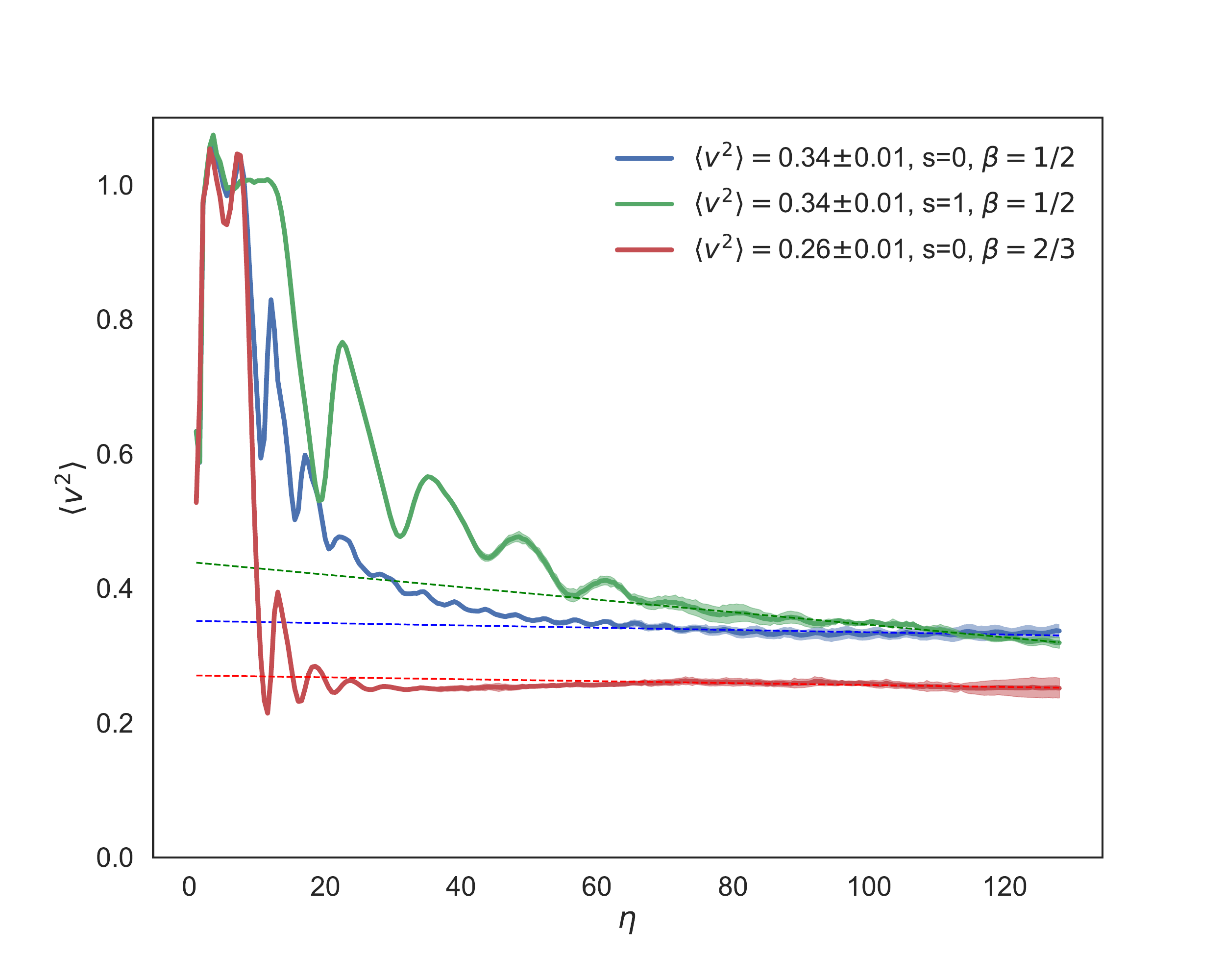}\includegraphics[width=8cm]{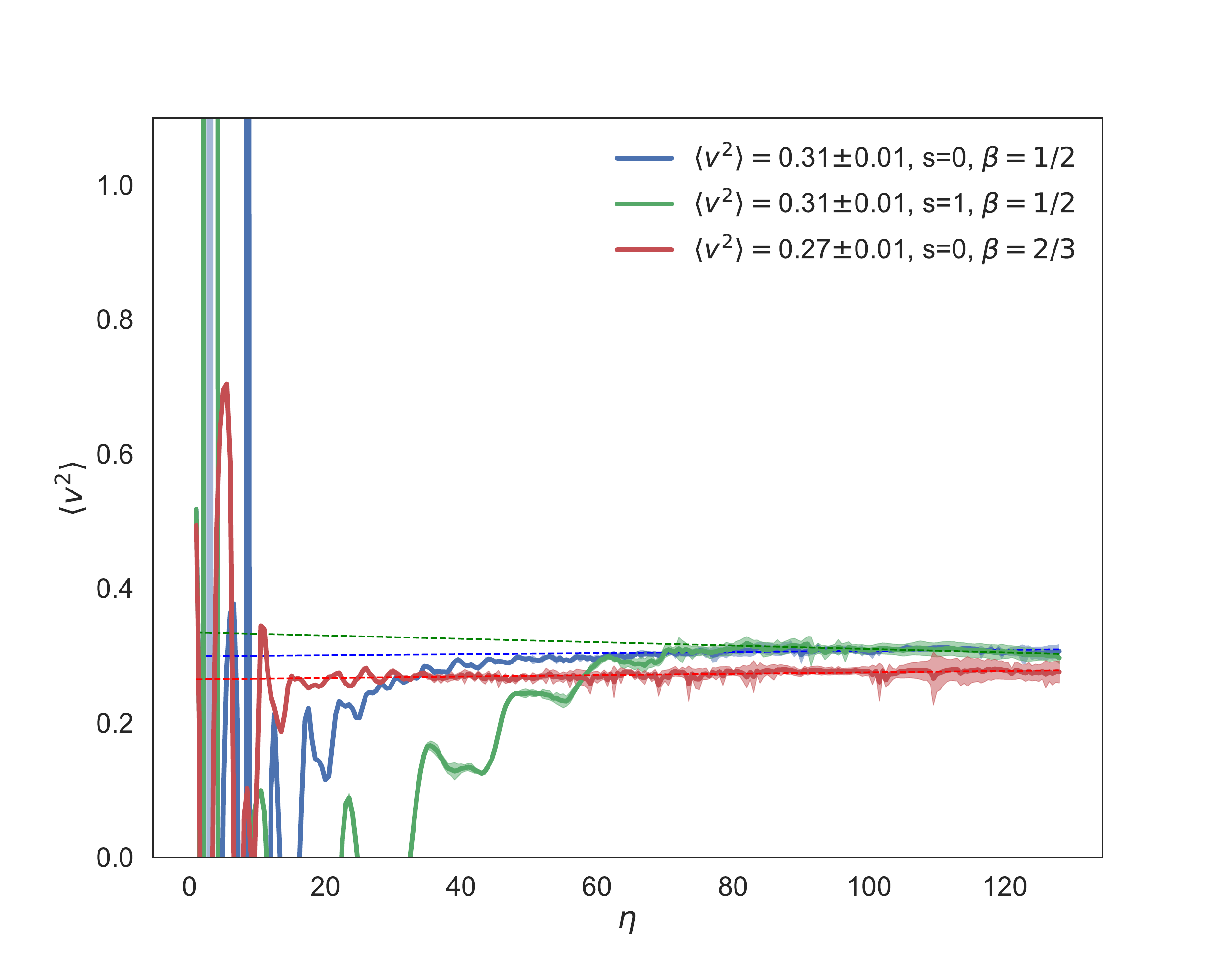}
\includegraphics[width=8cm]{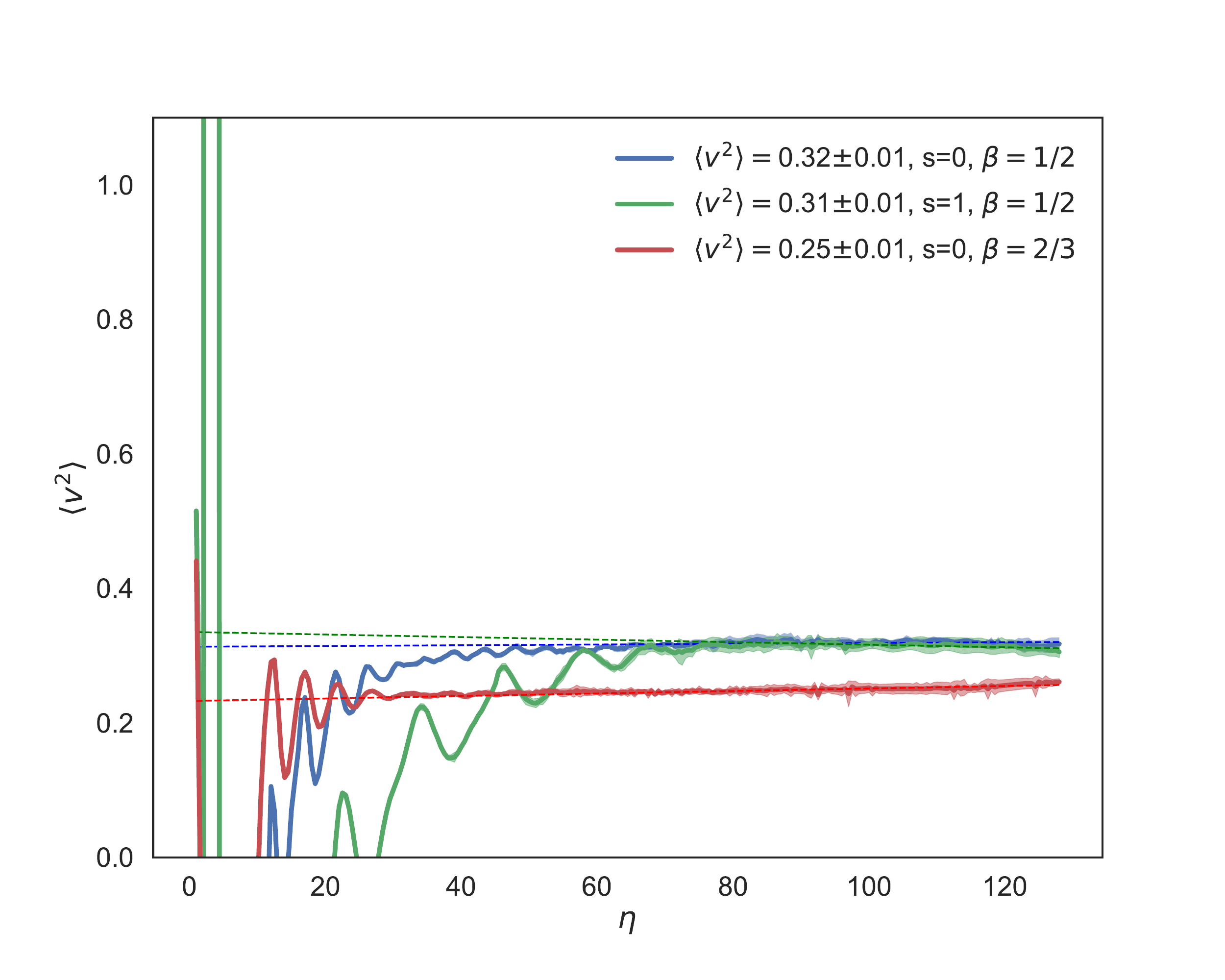}
\caption{\label{fig4}The evolution of the mean square velocity, estimated in three different ways: by using the estimator of Eq. \ref{defXiV2} weighted by the potential $\langle v^2 \rangle_{\mathcal{V}}$ (top left panel) or the Lagrangian  $\langle v^2 \rangle_{\mathcal{L}}$ (top right panel), or by using the equation of state parameter $\langle v^2 \rangle_{\omega}$ (bottom panel). In all cases the results are from $512^3$ runs, in the radiation era ($\beta=1/2$, blue lines without core growth and green lines with core growth) and in the matter era ($\beta=2/3$ red lines without core growth). The asymptotic values of the velocities, inferred after the networks have reached scaling, are also depicted.}
\end{center}
\end{figure*}

That being said, our analysis shows that all velocity estimators are in reasonable agreement. For radiation the potential-weighted estimator yields a slightly higher value than the others, but this is not statistically significant. Note that in the matter era we cannot evolve the true equations of motion, i.e. the case $s=1$ (one needs a larger dynamic range in order to successfully use core-growth), though this can be resolved by running larger simulation boxes as discussed in the next section. As such, we have compared our $s=0$ case to the $s=1$ value of \citet{Hindmarsh:2017qff}, for matter velocities; previous work, including that of \citet{Hindmarsh:2017qff} itself, suggests that this is not a significant issue. Plots of all the velocity diagnostics throughout the duration of the simulation in both radiation and matter era (with and without core growth, where applicable) can be found in Figure \ref{fig4}. As expected there are very large oscillations at early times (which the network relaxes from the choice of numerical initial conditions), but the approach to the constant-velocity scaling solution is clear at late times.


\section{\label{concl}Conclusions and outlook}

We have implemented field theory cosmic string evolution for the $U(1)$ model using the Compute Unified Device Architecture, such that it uses Graphics Processing Units as accelerators. We summarised the main implementation steps in terms of the performance of each kernel and showcased the achievable performance. In addition we compared the key physical diagnostic parameters for the mean string separation and the mean velocity squared to those previously reported in the literature, finding very good agreement and thus providing a preliminary validation of the code.

Compared to our previous GPGPU application---GPUwalls, see \citet{PhysRevE.96.043310}---the main bottlenecks in the present one are the evolution kernels, since one can force the calculation of useful quantities to occur every few timesteps and not every single timestep. This means that in contrast to the previous code we evade being compute-bound completely. The walls code also does not use software pre-fetching as was used here. Implementing these strategies is a task left for subsequent work.

The main challenges regarding the scalability of this code lie not only in being memory-bound but also in its memory requirements: given that two vector fields (float4's) and two complex scalar fields (float2's) are stored in $48$ bytes per lattice site, the largest box one could possibly simulate with one GPU is, at the time of writing, $512^3$ (the largest GPU memory in a commercial GPU is around $16$~GB). This brings us to our next step: to extend this simulation with multi-GPU support. In principle, given the large necessary number of GPU's required, the most natural way to implement multi-node, multi-GPU support would be through the Message-Passing-Interface. Note that in the multi-GPU case, the two main performance bottlenecks will be the presence of communications at every timestep (which can limit weak and strong scaling) or a small box size per GPU (where not enough threads are spawned to hide latency successfully, the expected main limiter of strong scaling). In the future we will report on the performance of such a version and show that near perfect weak scaling to thousands of GPU's can be achieved as long as one hides the communication cost properly, ie. by overlapping compute with communication at every timestep. For now there is no easy solution to the second bottleneck, and, as such strong scaling is less than ideal. In any case, we currently do not foresee any impediment to doing $8192^3$ simulations in existing high-performance computing facilities.

Overall we conclude that there is a tangible performance benefit to using GPUs in field theory defect simulations, enabling the possibility of running thousands or tens of thousands of high-resolution field theory simulations of Abelian-Higgs strings in acceptable amounts of time. This opens several interesting possibilities for the further exploration of the cosmological consequences of these networks. For the future, the continued increase in memory bandwidth of global memory of GPUs \citep{CPguide} is promising for our application, as it is memory bound. Even in the case of existing GPU's, there are cards with higher memory bandwidth (such as the case of the Tesla P100, with $732 GB/s$).

In the short term, it will be possible to provide a more quantitative calibration of the velocity-dependent one-scale model \citep{MS1,MS2,Book}, as was recently done for domain walls \citep{Rybak1,Rybak2}; early results can already be found in  \citet{Correia:2019bdl}. The comparative analysis of the evolution of field theory cosmic string and domain wall networks is itself interesting, since one expects that different energy loss mechanisms (specifically loop or blob production and scalar radiation) play different roles. A calibration with maximum box sizes of $512^3$ in the relativistic regime has been performed by the authors \citep{Correia:2019bdl}.

We also note that a long-term open issue in the cosmic strings literature is understanding the different results obtained in Goto-Nambu simulations, for which there are several independent codes \citep{BB,AS,FRAC,VVO,Blanco}, and in field theory simulations, for which all recent results ultimately stem from a single code \citep{Bevis:2006mj,Hindmarsh:2017qff}. To the extent that comparisons can already be made, our results are consistent with these. The availability of an improved (better calibrated) velocity-dependent one-scale model can also enable a more detailed comparison between the results of the two types of codes.

In the longer term, an optimised multi-GPU can be used to yield thousands of accurate full-sky maps of cosmic microwave or gravitational wave backgrounds which can be used in the data analysis of forthcoming experiments, such as CORE or LISA. This will eliminate the current bottleneck in this analysis (so far one can only generate a few full-sky maps, or many maps of small sky patches) thus leading to more robust as well as more stringent constraints. In conclusion, we expect that GPU-based defect codes will in the medium term become the gold standard in the field. 

\section*{Acknowledgements}
This work was financed by FEDER---Fundo Europeu de Desenvolvimento Regional funds through the COMPETE 2020---Operational Programme for Competitiveness and Internationalisation (POCI), and by Portuguese funds through FCT - Funda\c c\~ao para a Ci\^encia e a Tecnologia in the framework of the project POCI-01-0145-FEDER-028987. J.R.C. is supported by an FCT fellowship (grant reference SFRH/BD/130445/2017). We gratefully acknowledge the support of NVIDIA Corporation with the donation of the Quadro P5000 GPU used for this research.

\bibliography{artigo}

\begin{thebibliography}{46}
\expandafter\ifx\csname natexlab\endcsname\relax\def\natexlab#1{#1}\fi
\providecommand{\url}[1]{\texttt{#1}}
\providecommand{\href}[2]{#2}
\providecommand{\path}[1]{#1}
\providecommand{\DOIprefix}{doi:}
\providecommand{\ArXivprefix}{arXiv:}
\providecommand{\URLprefix}{URL: }
\providecommand{\Pubmedprefix}{pmid:}
\providecommand{\doi}[1]{\href{http://dx.doi.org/#1}{\path{#1}}}
\providecommand{\Pubmed}[1]{\href{pmid:#1}{\path{#1}}}
\providecommand{\bibinfo}[2]{#2}
\ifx\xfnm\relax \def\xfnm[#1]{\unskip,\space#1}\fi
\bibitem[{Abbott et~al.(2018)}]{LIGODefects}
\bibinfo{author}{Abbott, B.P.}, et~al. (\bibinfo{collaboration}{Virgo, LIGO
  Scientific}), \bibinfo{year}{2018}.
\newblock \bibinfo{title}{{Constraints on cosmic strings using data from the
  first Advanced LIGO observing run}}.
\newblock \bibinfo{journal}{Phys. Rev.} \bibinfo{volume}{D97},
  \bibinfo{pages}{102002}.
\newblock \DOIprefix\doi{10.1103/PhysRevD.97.102002},
  \href{http://arxiv.org/abs/1712.01168}{\tt arXiv:1712.01168}.
\bibitem[{Achucarro et~al.(2014)Achucarro, Avgoustidis, Leite, Lopez-Eiguren,
  Martins, Nunes and Urrestilla}]{Semilocals}
\bibinfo{author}{Achucarro, A.}, \bibinfo{author}{Avgoustidis, A.},
  \bibinfo{author}{Leite, A.M.M.}, \bibinfo{author}{Lopez-Eiguren, A.},
  \bibinfo{author}{Martins, C.J.A.P.}, \bibinfo{author}{Nunes, A.S.},
  \bibinfo{author}{Urrestilla, J.}, \bibinfo{year}{2014}.
\newblock \bibinfo{title}{{Evolution of semilocal string networks: Large-scale
  properties}}.
\newblock \bibinfo{journal}{Phys. Rev.} \bibinfo{volume}{D89},
  \bibinfo{pages}{063503}.
\newblock \DOIprefix\doi{10.1103/PhysRevD.89.063503},
  \href{http://arxiv.org/abs/1312.2123}{\tt arXiv:1312.2123}.
\bibitem[{Ade et~al.(2014)}]{PlanckDefects}
\bibinfo{author}{Ade, P.A.R.}, et~al. (\bibinfo{collaboration}{Planck}),
  \bibinfo{year}{2014}.
\newblock \bibinfo{title}{{Planck 2013 results. XXV. Searches for cosmic
  strings and other topological defects}}.
\newblock \bibinfo{journal}{Astron. Astrophys.} \bibinfo{volume}{571},
  \bibinfo{pages}{A25}.
\newblock \DOIprefix\doi{10.1051/0004-6361/201321621},
  \href{http://arxiv.org/abs/1303.5085}{\tt arXiv:1303.5085}.
\bibitem[{Allen and Shellard(1990)}]{AS}
\bibinfo{author}{Allen, B.}, \bibinfo{author}{Shellard, E.P.S.},
  \bibinfo{year}{1990}.
\newblock \bibinfo{title}{Cosmic string evolution: A numerical simulation}.
\newblock \bibinfo{journal}{Phys. Rev. Lett.} \bibinfo{volume}{64},
  \bibinfo{pages}{119--122}.
\bibitem[{Bennett and Bouchet(1990)}]{BB}
\bibinfo{author}{Bennett, D.P.}, \bibinfo{author}{Bouchet, F.R.},
  \bibinfo{year}{1990}.
\newblock \bibinfo{title}{High resolution simulations of cosmic string
  evolution. 1. network evolution}.
\newblock \bibinfo{journal}{Phys. Rev.} \bibinfo{volume}{D41},
  \bibinfo{pages}{2408}.
\bibitem[{Bevis et~al.(2007)Bevis, Hindmarsh, Kunz and
  Urrestilla}]{Bevis:2006mj}
\bibinfo{author}{Bevis, N.}, \bibinfo{author}{Hindmarsh, M.},
  \bibinfo{author}{Kunz, M.}, \bibinfo{author}{Urrestilla, J.},
  \bibinfo{year}{2007}.
\newblock \bibinfo{title}{{CMB power spectrum contribution from cosmic strings
  using field-evolution simulations of the Abelian Higgs model}}.
\newblock \bibinfo{journal}{Phys. Rev.} \bibinfo{volume}{D75},
  \bibinfo{pages}{065015}.
\newblock \DOIprefix\doi{10.1103/PhysRevD.75.065015},
  \href{http://arxiv.org/abs/astro-ph/0605018}{\tt arXiv:astro-ph/0605018}.
\bibitem[{Bevis et~al.(2010)Bevis, Hindmarsh, Kunz and
  Urrestilla}]{Bevis:2010gj}
\bibinfo{author}{Bevis, N.}, \bibinfo{author}{Hindmarsh, M.},
  \bibinfo{author}{Kunz, M.}, \bibinfo{author}{Urrestilla, J.},
  \bibinfo{year}{2010}.
\newblock \bibinfo{title}{{CMB power spectra from cosmic strings: predictions
  for the Planck satellite and beyond}}.
\newblock \bibinfo{journal}{Phys. Rev.} \bibinfo{volume}{D82},
  \bibinfo{pages}{065004}.
\newblock \DOIprefix\doi{10.1103/PhysRevD.82.065004},
  \href{http://arxiv.org/abs/1005.2663}{\tt arXiv:1005.2663}.
\bibitem[{Bevis and Saffin(2008)}]{Bevis:2008hg}
\bibinfo{author}{Bevis, N.}, \bibinfo{author}{Saffin, P.M.},
  \bibinfo{year}{2008}.
\newblock \bibinfo{title}{{Cosmic string Y-junctions: A Comparison between
  field theoretic and Nambu-Goto dynamics}}.
\newblock \bibinfo{journal}{Phys. Rev.} \bibinfo{volume}{D78},
  \bibinfo{pages}{023503}.
\newblock \DOIprefix\doi{10.1103/PhysRevD.78.023503},
  \href{http://arxiv.org/abs/0804.0200}{\tt arXiv:0804.0200}.
\bibitem[{Binetruy et~al.(2012)Binetruy, Bohe, Caprini and Dufaux}]{LISA}
\bibinfo{author}{Binetruy, P.}, \bibinfo{author}{Bohe, A.},
  \bibinfo{author}{Caprini, C.}, \bibinfo{author}{Dufaux, J.F.},
  \bibinfo{year}{2012}.
\newblock \bibinfo{title}{{Cosmological Backgrounds of Gravitational Waves and
  eLISA/NGO: Phase Transitions, Cosmic Strings and Other Sources}}.
\newblock \bibinfo{journal}{JCAP} \bibinfo{volume}{1206}, \bibinfo{pages}{027}.
\newblock \DOIprefix\doi{10.1088/1475-7516/2012/06/027},
  \href{http://arxiv.org/abs/1201.0983}{\tt arXiv:1201.0983}.
\bibitem[{Blanco-Pillado et~al.(2011)Blanco-Pillado, Olum and Shlaer}]{Blanco}
\bibinfo{author}{Blanco-Pillado, J.J.}, \bibinfo{author}{Olum, K.D.},
  \bibinfo{author}{Shlaer, B.}, \bibinfo{year}{2011}.
\newblock \bibinfo{title}{{Large parallel cosmic string simulations: New
  results o n loop production}}.
\newblock \bibinfo{journal}{Phys. Rev.} \bibinfo{volume}{D83},
  \bibinfo{pages}{083514}.
\newblock \DOIprefix\doi{10.1103/PhysRevD.83.083514},
  \href{http://arxiv.org/abs/1101.5173}{\tt arXiv:1101.5173}.
\bibitem[{Briggs et~al.(2014)Briggs, Pennycook, Shellard, Martins, Woodacre and
  Feind}]{Intel}
\bibinfo{author}{Briggs, J.}, \bibinfo{author}{Pennycook, S.J.},
  \bibinfo{author}{Shellard, E.P.S.}, \bibinfo{author}{Martins, C.J.A.P.},
  \bibinfo{author}{Woodacre, M.}, \bibinfo{author}{Feind, K.},
  \bibinfo{year}{2014}.
\newblock \bibinfo{title}{{Unveiling the Early Universe: Optimizing Cosmology
  Workloads for Intel Xeon Phi Coprocessors in an SGI UV20 00 System}}.
\newblock \bibinfo{type}{Technical Report}. SGI/Intel White Paper.
\bibitem[{Correia and Martins(2017)}]{PhysRevE.96.043310}
\bibinfo{author}{Correia, J.R.C.C.C.}, \bibinfo{author}{Martins, C.J.A.P.},
  \bibinfo{year}{2017}.
\newblock \bibinfo{title}{General purpose graphics-processing-unit
  implementation of cosmological domain wall network evolution}.
\newblock \bibinfo{journal}{Phys. Rev. E} \bibinfo{volume}{96},
  \bibinfo{pages}{043310}.
\newblock \URLprefix \url{https://link.aps.org/doi/10.1103/PhysRevE.96.043310},
  \DOIprefix\doi{10.1103/PhysRevE.96.043310}.
\bibitem[{Correia and Martins(2019)}]{Correia:2019bdl}
\bibinfo{author}{Correia, J.R.C.C.C.}, \bibinfo{author}{Martins, J.A.P.},
  \bibinfo{year}{2019}.
\newblock \bibinfo{title}{{Extending and Calibrating the Velocity dependent
  One-Scale model for Cosmic Strings with One Thousand Field Theory
  Simulations}}.
\newblock \bibinfo{journal}{Phys. Rev.} \bibinfo{volume}{D100},
  \bibinfo{pages}{103517}.
\newblock \DOIprefix\doi{10.1103/PhysRevD.100.103517},
  \href{http://arxiv.org/abs/1911.03163}{\tt arXiv:1911.03163}.
\bibitem[{Daverio et~al.(2016)Daverio, Hindmarsh, Kunz, Lizarraga and
  Urrestilla}]{Daverio:2015nva}
\bibinfo{author}{Daverio, D.}, \bibinfo{author}{Hindmarsh, M.},
  \bibinfo{author}{Kunz, M.}, \bibinfo{author}{Lizarraga, J.},
  \bibinfo{author}{Urrestilla, J.}, \bibinfo{year}{2016}.
\newblock \bibinfo{title}{{Energy-momentum correlations for Abelian Higgs
  cosmic strings}}.
\newblock \bibinfo{journal}{Phys. Rev.} \bibinfo{volume}{D93},
  \bibinfo{pages}{085014}.
\newblock \DOIprefix\doi{10.1103/PhysRevD.95.049903,
  10.1103/PhysRevD.93.085014}, \href{http://arxiv.org/abs/1510.05006}{\tt
  arXiv:1510.05006}. \bibinfo{note}{[Erratum: Phys.
  Rev.D95,no.4,049903(2017)]}.
\bibitem[{Drew and Shellard(2019)}]{GlobalD}
\bibinfo{author}{Drew, A.}, \bibinfo{author}{Shellard, E.P.S.},
  \bibinfo{year}{2019}.
\newblock \bibinfo{title}{{Radiation from Global Topological Strings using
  Adaptive Mesh Refinement: Methodology and Massless Modes}}
  \href{http://arxiv.org/abs/1910.01718}{\tt arXiv:1910.01718}.
\bibitem[{Finelli et~al.(2018)}]{CORE}
\bibinfo{author}{Finelli, F.}, et~al. (\bibinfo{collaboration}{CORE}),
  \bibinfo{year}{2018}.
\newblock \bibinfo{title}{{Exploring cosmic origins with CORE: Inflation}}.
\newblock \bibinfo{journal}{JCAP} \bibinfo{volume}{1804}, \bibinfo{pages}{016}.
\newblock \DOIprefix\doi{10.1088/1475-7516/2018/04/016},
  \href{http://arxiv.org/abs/1612.08270}{\tt arXiv:1612.08270}.
\bibitem[{Helfer et~al.(2019)Helfer, Aurrekoetxea and Lim}]{Helfer}
\bibinfo{author}{Helfer, T.}, \bibinfo{author}{Aurrekoetxea, J.C.},
  \bibinfo{author}{Lim, E.A.}, \bibinfo{year}{2019}.
\newblock \bibinfo{title}{{Cosmic String Loop Collapse in Full General
  Relativity}}.
\newblock \bibinfo{journal}{Phys. Rev.} \bibinfo{volume}{D99},
  \bibinfo{pages}{104028}.
\newblock \DOIprefix\doi{10.1103/PhysRevD.99.104028},
  \href{http://arxiv.org/abs/1808.06678}{\tt arXiv:1808.06678}.
\bibitem[{Hindmarsh and Daverio()}]{Hind:email}
\bibinfo{author}{Hindmarsh, M.}, \bibinfo{author}{Daverio, D.}, .
\newblock \bibinfo{howpublished}{Private communication, 20 December 2019}.
\bibitem[{Hindmarsh et~al.(2018)Hindmarsh, Kormu, Lopez-Eiguren and
  Weir}]{Hindmarsh:2018zch}
\bibinfo{author}{Hindmarsh, M.}, \bibinfo{author}{Kormu, A.},
  \bibinfo{author}{Lopez-Eiguren, A.}, \bibinfo{author}{Weir, D.J.},
  \bibinfo{year}{2018}.
\newblock \bibinfo{title}{{Scaling in necklaces of monopoles and semipoles}}.
\newblock \bibinfo{journal}{Phys. Rev.} \bibinfo{volume}{D98},
  \bibinfo{pages}{103533}.
\newblock \DOIprefix\doi{10.1103/PhysRevD.98.103533},
  \href{http://arxiv.org/abs/1809.03384}{\tt arXiv:1809.03384}.
\bibitem[{Hindmarsh et~al.(2020)Hindmarsh, Lizarraga, Lopez-Eiguren and
  Urrestilla}]{GlobalH}
\bibinfo{author}{Hindmarsh, M.}, \bibinfo{author}{Lizarraga, J.},
  \bibinfo{author}{Lopez-Eiguren, A.}, \bibinfo{author}{Urrestilla, J.},
  \bibinfo{year}{2020}.
\newblock \bibinfo{title}{{The scaling density of axion strings}}.
\newblock \bibinfo{journal}{Phys. Rev. Lett.} \bibinfo{volume}{124},
  \bibinfo{pages}{021301}.
\newblock \DOIprefix\doi{10.1103/PhysRevLett.124.021301},
  \href{http://arxiv.org/abs/1908.03522}{\tt arXiv:1908.03522}.
\bibitem[{Hindmarsh et~al.(2017a)Hindmarsh, Lizarraga, Urrestilla, Daverio and
  Kunz}]{Hindmarsh:2017qff}
\bibinfo{author}{Hindmarsh, M.}, \bibinfo{author}{Lizarraga, J.},
  \bibinfo{author}{Urrestilla, J.}, \bibinfo{author}{Daverio, D.},
  \bibinfo{author}{Kunz, M.}, \bibinfo{year}{2017}a.
\newblock \bibinfo{title}{{Scaling from gauge and scalar radiation in Abelian
  Higgs string networks}}.
\newblock \bibinfo{journal}{Phys. Rev.} \bibinfo{volume}{D96},
  \bibinfo{pages}{023525}.
\newblock \DOIprefix\doi{10.1103/PhysRevD.96.023525},
  \href{http://arxiv.org/abs/1703.06696}{\tt arXiv:1703.06696}.
\bibitem[{Hindmarsh et~al.(2019)Hindmarsh, Lizarraga, Urrestilla, Daverio and
  Kunz}]{Type1}
\bibinfo{author}{Hindmarsh, M.}, \bibinfo{author}{Lizarraga, J.},
  \bibinfo{author}{Urrestilla, J.}, \bibinfo{author}{Daverio, D.},
  \bibinfo{author}{Kunz, M.}, \bibinfo{year}{2019}.
\newblock \bibinfo{title}{{Type I Abelian Higgs strings: evolution and Cosmic
  Microwave Background constraints}}.
\newblock \bibinfo{journal}{Phys. Rev.} \bibinfo{volume}{D99},
  \bibinfo{pages}{083522}.
\newblock \DOIprefix\doi{10.1103/PhysRevD.99.083522},
  \href{http://arxiv.org/abs/1812.08649}{\tt arXiv:1812.08649}.
\bibitem[{Hindmarsh et~al.(2017b)Hindmarsh, Rummukainen and Weir}]{HindNab}
\bibinfo{author}{Hindmarsh, M.}, \bibinfo{author}{Rummukainen, K.},
  \bibinfo{author}{Weir, D.J.}, \bibinfo{year}{2017}b.
\newblock \bibinfo{title}{{Numerical simulations of necklaces in SU(2)
  gauge-Higgs field theory}}.
\newblock \bibinfo{journal}{Phys. Rev.} \bibinfo{volume}{D95},
  \bibinfo{pages}{063520}.
\newblock \DOIprefix\doi{10.1103/PhysRevD.95.063520},
  \href{http://arxiv.org/abs/1611.08456}{\tt arXiv:1611.08456}.
\bibitem[{Kajantie et~al.(1998)Kajantie, Karjalainen, Laine, Peisa and
  Rajantie}]{Kajantie:1998bg}
\bibinfo{author}{Kajantie, K.}, \bibinfo{author}{Karjalainen, M.},
  \bibinfo{author}{Laine, M.}, \bibinfo{author}{Peisa, J.},
  \bibinfo{author}{Rajantie, A.}, \bibinfo{year}{1998}.
\newblock \bibinfo{title}{{Thermodynamics of gauge invariant U(1) vortices from
  lattice Monte Carlo simulations}}.
\newblock \bibinfo{journal}{Phys. Lett.} \bibinfo{volume}{B428},
  \bibinfo{pages}{334--341}.
\newblock \DOIprefix\doi{10.1016/S0370-2693(98)00440-7},
  \href{http://arxiv.org/abs/hep-ph/9803367}{\tt arXiv:hep-ph/9803367}.
\bibitem[{Kibble(1976)}]{Kibble:1976sj}
\bibinfo{author}{Kibble, T.W.B.}, \bibinfo{year}{1976}.
\newblock \bibinfo{title}{{Topology of Cosmic Domains and Strings}}.
\newblock \bibinfo{journal}{J. Phys.} \bibinfo{volume}{A9},
  \bibinfo{pages}{1387--1398}.
\newblock \DOIprefix\doi{10.1088/0305-4470/9/8/029}.
\bibitem[{Lizarraga and Urrestilla(2016)}]{Lizarraga:2016hpd}
\bibinfo{author}{Lizarraga, J.}, \bibinfo{author}{Urrestilla, J.},
  \bibinfo{year}{2016}.
\newblock \bibinfo{title}{{Survival of pq-superstrings in field theory
  simulations}}.
\newblock \bibinfo{journal}{JCAP} \bibinfo{volume}{1604}, \bibinfo{pages}{053}.
\newblock \DOIprefix\doi{10.1088/1475-7516/2016/04/053},
  \href{http://arxiv.org/abs/1602.08014}{\tt arXiv:1602.08014}.
\bibitem[{Lopez-Eiguren et~al.(2017)Lopez-Eiguren, Urrestilla and
  Achucarro}]{Monopoles}
\bibinfo{author}{Lopez-Eiguren, A.}, \bibinfo{author}{Urrestilla, J.},
  \bibinfo{author}{Achucarro, A.}, \bibinfo{year}{2017}.
\newblock \bibinfo{title}{{Measuring Global Monopole Velocities, one by one}}.
\newblock \bibinfo{journal}{JCAP} \bibinfo{volume}{1701}, \bibinfo{pages}{020}.
\newblock \DOIprefix\doi{10.1088/1475-7516/2017/01/020},
  \href{http://arxiv.org/abs/1611.09628}{\tt arXiv:1611.09628}.
\bibitem[{Martins(2016)}]{Book}
\bibinfo{author}{Martins, C.J.A.P.}, \bibinfo{year}{2016}.
\newblock \bibinfo{title}{Defect Evolution in Cosmology and Condensed Matter:
  Quantitative Analysis with the Velocity-Dependent One-Scale Model}.
\newblock \bibinfo{publisher}{Springer}.
\bibitem[{Martins et~al.(2016a)Martins, Rybak, Avgoustidis and
  Shellard}]{Rybak1}
\bibinfo{author}{Martins, C.J.A.P.}, \bibinfo{author}{Rybak, I.Y.},
  \bibinfo{author}{Avgoustidis, A.}, \bibinfo{author}{Shellard, E.P.S.},
  \bibinfo{year}{2016}a.
\newblock \bibinfo{title}{{Extending the velocity-dependent one-scale model for
  domain walls}}.
\newblock \bibinfo{journal}{Phys. Rev.} \bibinfo{volume}{D93},
  \bibinfo{pages}{043534}.
\newblock \DOIprefix\doi{10.1103/PhysRevD.93.043534},
  \href{http://arxiv.org/abs/1602.01322}{\tt arXiv:1602.01322}.
\bibitem[{Martins et~al.(2016b)Martins, Rybak, Avgoustidis and
  Shellard}]{Rybak2}
\bibinfo{author}{Martins, C.J.A.P.}, \bibinfo{author}{Rybak, I.{\relax Yu}.},
  \bibinfo{author}{Avgoustidis, A.}, \bibinfo{author}{Shellard, E.P.S.},
  \bibinfo{year}{2016}b.
\newblock \bibinfo{title}{{Stretching and Kibble scaling regimes for
  Hubble-damped defect networks}}.
\newblock \bibinfo{journal}{Phys. Rev.} \bibinfo{volume}{D94},
  \bibinfo{pages}{116017}.
\newblock \DOIprefix\doi{10.1103/PhysRevD.94.116017,
  10.1103/PhysRevD.95.039902}, \href{http://arxiv.org/abs/1612.08863}{\tt
  arXiv:1612.08863}. \bibinfo{note}{[Erratum: Phys.
  Rev.D95,no.3,039902(2017)]}.
\bibitem[{Martins and Shellard(1996)}]{MS1}
\bibinfo{author}{Martins, C.J.A.P.}, \bibinfo{author}{Shellard, E.P.S.},
  \bibinfo{year}{1996}.
\newblock \bibinfo{title}{Quantitative string evolution}.
\newblock \bibinfo{journal}{Phys. Rev.} \bibinfo{volume}{D54},
  \bibinfo{pages}{2535--2556}.
\newblock \href{http://arxiv.org/abs/hep-ph/9602271}{\tt arXiv:hep-ph/9602271}.
\bibitem[{Martins and Shellard(2002)}]{MS2}
\bibinfo{author}{Martins, C.J.A.P.}, \bibinfo{author}{Shellard, E.P.S.},
  \bibinfo{year}{2002}.
\newblock \bibinfo{title}{Extending the velocity-dependent one-scale string
  evolution model}.
\newblock \bibinfo{journal}{Phys. Rev.} \bibinfo{volume}{D65},
  \bibinfo{pages}{043514}.
\newblock \href{http://arxiv.org/abs/hep-ph/0003298}{\tt arXiv:hep-ph/0003298}.
\bibitem[{Martins and Shellard(2006)}]{FRAC}
\bibinfo{author}{Martins, C.J.A.P.}, \bibinfo{author}{Shellard, E.P.S.},
  \bibinfo{year}{2006}.
\newblock \bibinfo{title}{Fractal properties and small-scale structure of
  cosmic string network s}.
\newblock \bibinfo{journal}{Phys. Rev.} \bibinfo{volume}{D73},
  \bibinfo{pages}{043515}.
\newblock \href{http://arxiv.org/abs/astro-ph/0511792}{\tt
  arXiv:astro-ph/0511792}.
\bibitem[{McGraw(1998)}]{McGraw}
\bibinfo{author}{McGraw, P.}, \bibinfo{year}{1998}.
\newblock \bibinfo{title}{{Evolution of a nonAbelian cosmic string network}}.
\newblock \bibinfo{journal}{Phys. Rev.} \bibinfo{volume}{D57},
  \bibinfo{pages}{3317--3339}.
\newblock \DOIprefix\doi{10.1103/PhysRevD.57.3317},
  \href{http://arxiv.org/abs/astro-ph/9706182}{\tt arXiv:astro-ph/9706182}.
\bibitem[{Micikevicius(2009)}]{Micikevicius}
\bibinfo{author}{Micikevicius, P.}, \bibinfo{year}{2009}.
\newblock \bibinfo{title}{3d finite difference computation on gpus using cuda},
  in: \bibinfo{booktitle}{Proceedings of 2Nd Workshop on General Purpose
  Processing on Graphics Processing Units}, \bibinfo{publisher}{ACM},
  \bibinfo{address}{New York, NY, USA}. pp. \bibinfo{pages}{79--84}.
\newblock \URLprefix \url{http://doi.acm.org/10.1145/1513895.1513905},
  \DOIprefix\doi{10.1145/1513895.1513905}.
\bibitem[{Nguyen et~al.(2010)Nguyen, Satish, Chhugani, Kim and Dubey}]{5645463}
\bibinfo{author}{Nguyen, A.}, \bibinfo{author}{Satish, N.},
  \bibinfo{author}{Chhugani, J.}, \bibinfo{author}{Kim, C.},
  \bibinfo{author}{Dubey, P.}, \bibinfo{year}{2010}.
\newblock \bibinfo{title}{3.5-d blocking optimization for stencil computations
  on modern cpus and gpus}, in: \bibinfo{booktitle}{2010 ACM/IEEE International
  Conference for High Performance Computing, Networking, Storage and Analysis},
  pp. \bibinfo{pages}{1--13}.
\newblock \DOIprefix\doi{10.1109/SC.2010.2}.
\bibitem[{NvidiaCorporation(a)}]{CPguide}
\bibinfo{author}{NvidiaCorporation}, a.
\newblock \bibinfo{title}{Cuda programming guide}.
\newblock
  \bibinfo{howpublished}{\url{https://docs.nvidia.com/cuda/cuda-c-programming-guide/index.html}}.
\bibitem[{NvidiaCorporation(b)}]{curand}
\bibinfo{author}{NvidiaCorporation}, b.
\newblock \bibinfo{title}{curand}.
\newblock \bibinfo{howpublished}{\url{https://developer.nvidia.com/curand}}.
\bibitem[{NvidiaResearch-NVLabs(2018)}]{CUB}
\bibinfo{author}{NvidiaResearch-NVLabs}, \bibinfo{year}{2018}.
\newblock \bibinfo{title}{Cub - cuda unbound v1.8.0}.
\newblock \bibinfo{howpublished}{\url{https://nvlabs.github.io/cub/}}.
\bibitem[{Olum and Vanchurin(2007)}]{VVO}
\bibinfo{author}{Olum, K.D.}, \bibinfo{author}{Vanchurin, V.},
  \bibinfo{year}{2007}.
\newblock \bibinfo{title}{Cosmic string loops in the expanding universe}.
\newblock \bibinfo{journal}{Phys. Rev.} \bibinfo{volume}{D75},
  \bibinfo{pages}{063521}.
\newblock \href{http://arxiv.org/abs/astro-ph/0610419}{\tt
  arXiv:astro-ph/0610419}.
\bibitem[{Phillips and Fatica(2010)}]{5470394}
\bibinfo{author}{Phillips, E.H.}, \bibinfo{author}{Fatica, M.},
  \bibinfo{year}{2010}.
\newblock \bibinfo{title}{Implementing the himeno benchmark with cuda on gpu
  clusters}, in: \bibinfo{booktitle}{2010 IEEE International Symposium on
  Parallel Distributed Processing (IPDPS)}, pp. \bibinfo{pages}{1--10}.
\newblock \DOIprefix\doi{10.1109/IPDPS.2010.5470394}.
\bibitem[{Press et~al.(1989)Press, Ryden and Spergel}]{PRS}
\bibinfo{author}{Press, W.H.}, \bibinfo{author}{Ryden, B.S.},
  \bibinfo{author}{Spergel, D.N.}, \bibinfo{year}{1989}.
\newblock \bibinfo{title}{{Dynamical Evolution of Domain Walls in an Expanding
  Universe}}.
\newblock \bibinfo{journal}{Astrophys. J.} \bibinfo{volume}{347},
  \bibinfo{pages}{590--604}.
\newblock \DOIprefix\doi{10.1086/168151}.
\bibitem[{Scherrer and Vilenkin(1998)}]{PhysRevD.58.103501}
\bibinfo{author}{Scherrer, R.J.}, \bibinfo{author}{Vilenkin, A.},
  \bibinfo{year}{1998}.
\newblock \bibinfo{title}{{'Lattice-free' simulations of topological defect
  formation}}.
\newblock \bibinfo{journal}{Phys. Rev.} \bibinfo{volume}{D58},
  \bibinfo{pages}{103501}.
\newblock \DOIprefix\doi{10.1103/PhysRevD.58.103501},
  \href{http://arxiv.org/abs/hep-ph/9709498}{\tt arXiv:hep-ph/9709498}.
\bibitem[{Wilson(1974)}]{Wilson:1974sk}
\bibinfo{author}{Wilson, K.G.}, \bibinfo{year}{1974}.
\newblock \bibinfo{title}{{Confinement of Quarks}}.
\newblock \bibinfo{journal}{Phys. Rev.} \bibinfo{volume}{D10},
  \bibinfo{pages}{2445--2459}.
\newblock \DOIprefix\doi{10.1103/PhysRevD.10.2445}.
  \bibinfo{note}{[,319(1974)]}.
\bibitem[{Xmartlabs(2012)}]{occup}
\bibinfo{author}{Xmartlabs}, \bibinfo{year}{2012}.
\newblock \bibinfo{title}{Cuda occupancy calculator}.
\newblock
  \bibinfo{howpublished}{\url{https://github.com/xmartlabs/cuda-calculator}}.
\bibitem[{Zhang and Mueller(2012)}]{Zhang}
\bibinfo{author}{Zhang, Y.}, \bibinfo{author}{Mueller, F.},
  \bibinfo{year}{2012}.
\newblock \bibinfo{title}{Auto-generation and auto-tuning of 3d stencil codes
  on gpu clusters}, in: \bibinfo{booktitle}{Proceedings of the Tenth
  International Symposium on Code Generation and Optimization},
  \bibinfo{publisher}{ACM}, \bibinfo{address}{New York, NY, USA}. pp.
  \bibinfo{pages}{155--164}.
\newblock \URLprefix \url{http://doi.acm.org/10.1145/2259016.2259037},
  \DOIprefix\doi{10.1145/2259016.2259037}.

\end{thebibliography}
\end{document}